\pdfoutput=1
\documentclass[sigconf,nonacm]{acmart}
\usepackage{colortbl} 
\usepackage{tikz}
\usepackage{adjustbox}
\usetikzlibrary{shadows} 
\usepackage{balance}
\usepackage{tabularx}
\usepackage{array}     
\usepackage{makecell}   
\usepackage{booktabs}
\usepackage[T1]{fontenc}
\usepackage[normalem]{ulem} 
\usepackage{microtype}         

\usepackage{amsmath, amssymb, amsfonts}
\usepackage{xcolor}          
\usepackage{graphicx}        
\usepackage{dblfloatfix}      
\usepackage{caption}
\usepackage{subcaption}
\usepackage{float} 
\usepackage{algorithm}
\usepackage{algpseudocode}    
\algrenewcommand\algorithmiccomment[1]{\hfill\footnotesize\textcolor{gray}{// #1}}
\usepackage{listings}
\usepackage{multirow}
\usepackage{enumitem}         
\usepackage{placeins}

\newcommand{\rev}[1]{\textcolor{blue!80!black}{#1}}

\newcommand{\revm}[2]{\rev{\sout{#1} #2}}

\newif\ifshowchanges
\showchangesfalse     

\ifshowchanges
  \renewcommand{\rev}[1]{\textcolor{blue!80!black}{#1}}
  \renewcommand{\revm}[2]{\textcolor{blue!80!black}{\sout{#1} #2}}
\else
  \renewcommand{\rev}[1]{#1}
  \renewcommand{\revm}[2]{#2}
\fi

\AtBeginDocument{%
  \providecommand\BibTeX{{%
    \normalfont B\kern-0.5em{\scshape i\kern-0.25em b}\kern-0.8em\TeX}}}

\acmConference[ICSE '26]%
  {Proceedings of the 48th IEEE/ACM International Conference on Software Engineering (ICSE 2026)}%
  {April 12–18 2026}%
  {Rio de Janeiro, Brazil}

\begin{document}

\title{TAAF: A Trace Abstraction and Analysis Framework Synergizing Knowledge Graphs and LLMs}
\thanks{Accepted for publication at ICSE 2026. DOI: \href{https://doi.org/10.1145/3744916.3787832}{10.1145/3744916.3787832}.}



\author{Alireza Ezaz}
\orcid{0009-0001-4156-2750}
\affiliation{%
\institution{Brock University}
\department{Computer Science}
\streetaddress{1812 Sir Isaac Brock Way}
\city{St. Catharines}
\state{Ontario}
\postcode{L2S3A1}
\country{Canada}}
\email{sezaz@brocku.ca}

\author{Ghazal Khodabandeh}
\orcid{0009-0001-4587-1876}
\affiliation{%
\institution{Brock University}
\department{Computer Science}
\streetaddress{1812 Sir Isaac Brock Way}
\city{St. Catharines}
\state{Ontario}
\postcode{L2S3A1}
\country{Canada}}
\email{gkhodobandeh@brocku.ca}

\author{Majid Babaei}
\orcid{0000-0002-1394-4030}
\affiliation{%
\institution{Mcgil University}
\department{Computer Science}
\streetaddress{845 Sherbrooke St W, Montreal}
\city{Montreal}
\state{Quebec}
\postcode{H3A 0G4}
\country{Canada}}
\email{majid.babaei@mcgill.ca}

\author{Naser Ezzati-Jivan}
\orcid{0000-0003-1435-6297}
\affiliation{%
\institution{Brock University}
\department{Computer Science}
\streetaddress{1812 Sir Isaac Brock Way}
\city{St. Catharines}
\state{Ontario}
\postcode{L2S3A1}
\country{Canada}}
\email{nezzatijivan@brocku.ca}


\begin{abstract}
Execution traces are a critical source of information for understanding, debugging, and optimizing complex software systems. However, traces from OS kernels or large-scale applications like Chrome or MySQL are massive and difficult to analyze. Existing tools rely on predefined analyses, and custom insights often require writing domain-specific scripts, which is an error-prone and time-consuming task.
This paper introduces \textbf{TAAF} (Trace Abstraction and Analysis Framework), a novel approach that combines time-indexing, knowledge graphs (KGs), and large language models (LLMs) to transform raw trace data into actionable insights. TAAF constructs a time-indexed KG from trace events to capture relationships among entities such as threads, CPUs, and system resources. An LLM then interprets query-specific subgraphs to answer natural-language questions, reducing the need for manual inspection and deep system expertise.
To evaluate TAAF, we introduce \textbf{TraceQA-100}, a benchmark of 100 questions grounded in real kernel traces. Experiments across three LLMs and multiple temporal settings show that TAAF improves answer accuracy by up to 31.2\%, particularly in multi-hop and causal reasoning tasks. We further analyze where graph-grounded reasoning helps and where limitations remain, offering a foundation for next-generation trace analysis tools.
\end{abstract}




\maketitle

\section{Introduction}
\label{sec:introduction}

Execution traces are a foundational source of evidence for understanding, debugging, and optimizing software systems. They capture fine-grained temporal telemetry from across the stack, ranging from user-level function calls and inter-service messages to low-level kernel events such as system calls, context switches, and interrupt handling. Tools such as Zipkin~\cite{zipkin}, Jaeger~\cite{jaeger}, LTTng~\cite{lttng}, and Trace Compass~\cite{tracecompass} provide access to such data, but interpreting it at scale remains difficult. Effective analysis requires deep expertise in OS internals, hardware scheduling, or file system behavior. Existing tools offer limited flexibility. Users must rely on rigid charts or write custom scripts using low-level APIs, a process that is \revm{fragile,}{} error-prone, and inaccessible to non-experts.

Several challenges make trace analysis hard to scale. Traces are massive. Even short runs can produce hundreds of millions of events, which cannot fit in memory or in the context window of a large language model (LLM). They are also multidimensional. Events involve threads, processes, files, sockets, and CPUs interacting over time. Finally, reasoning often requires correlating events across long time intervals and interpreting raw numerical fields with little semantic guidance. Even simple questions like "Why was this file closed?" may demand deep manual effort. These limitations make trace analysis hard to democratize. \rev{To ground the problem we begin with a short operational scenario that reflects a common trace triage scene.}

\rev{
\begin{flushleft}
\begin{tikzpicture}
\node[
  draw=yellow!70!black,
  fill=yellow!8,
  thick,
  rounded corners=6pt,
  inner sep=7pt,
  anchor=west,
  text width=\dimexpr\linewidth-15pt\relax
] {
  \textbf{Motivating scenario.}
  A performance engineer investigates a recurring latency spike on CPU\_1  during a load test. Warm up has passed and the spike repeats within a short window. The first triage step is to decide whether CPU\_1 is dominated by one thread or by many. The engineer isolates the interval \([t_1, t_2]\) where the spike appears and asks:
  \begin{itemize}[leftmargin=*]
    \item Which thread executed most on CPU\_1 between \(t_1\) and \(t_2\)\,?
    \item Within the same window, how many distinct threads ran on CPU\_1 and did the top thread account for the majority of runtime\,?
  \end{itemize}
};
\end{tikzpicture}
\end{flushleft}
}

\rev{These checks are representative of common and time-consuming tasks because they expose imbalance and migration patterns that decide the next diagnostic step. Here teams follow playbooks that first identify the top consumer on the affected CPU then measure diversity to decide whether to pursue one thread or investigate system wide pressure. Both checks are time consuming when done by hand since the analyst must sum per thread run time over the exact interval and account for migrations. Prior work reports 15--45 minutes per investigation \cite{montplaisir2013efficient}. This is one example of such investigations. Other incidents may favor analogous interval scoped checks such as the same pair on CPU\_0 or the same procedure in a 10s window. 
}



\revm{We propose}{Building on this motivation, we propose} \textbf{TAAF} (\emph{Trace Abstraction and Analysis with Foundation models}), a three-layered framework for semantically meaningful and scalable trace analysis. Instead of feeding raw data into an LLM, TAAF builds a compact \revm{symbolic}{structured} abstraction. A temporal state system summarizes millions of events into structured, time-indexed transitions; a query-specific knowledge graph organizes entities and their relationships; and a large language model interprets the graph to answer natural-language questions. This modular design enables explainable reasoning, improves scalability, and reduces hallucinations in complex system scenarios.

Our research investigates how \revm{symbolic}{structured} abstractions like state systems and knowledge graphs can bridge low-level event data and high-level questions. Specifically, we ask: (1) How can knowledge graphs model the temporal and multidimensional nature of traces? (2) Can indexed trace outputs improve LLM reasoning over raw logs? (3) How does combining KGs with LLMs affect the accuracy and explainability of trace analysis? These questions guide the design and evaluation of TAAF.

\textbf{We make the following contributions:}

\begin{itemize}
\item We present \textbf{TAAF}, a three-layered framework combining temporal abstraction, knowledge graphs, and LLM reasoning for semantic trace analysis.
\item We propose a method to transform kernel traces into time-indexed graphs that encode entity interactions and temporal structure.
\item We introduce \textbf{TraceQA-100}, a benchmark of 100 expert-authored trace-analysis questions grounded in real Linux traces, focusing on thread–CPU interactions and activity patterns related to scheduling behavior.
\item We empirically show that graph-grounded reasoning improves answer accuracy (up to 31\%) across GPT-4o, GPT-4.1 nano and o4-mini (reasoning), relative to raw or flattened inputs.
\end{itemize}


We evaluate our approach using real Linux traces by segmenting the trace into varying temporal locations and durations to assess scalability. Our experiments involve multiple LLMs and grounding strategies. TAAF consistently outperforms the baselines, demonstrating that \revm{symbolic}{structured} grounding enhances both reasoning accuracy and robustness. We also anonymously release all code, data, and evaluation artifacts to support reproducibility and future research\footnote{\url{https://anonymous.4open.science/r/TAAF-LLM-KG-State-System--60EC/README.md}}.

\section{Related Work}
\label{sec:related-work}

We review prior work in four key areas relevant to TAAF: trace abstraction and stateful modeling, knowledge graph construction for system understanding, large language models for software reasoning, and integration strategies that bridge \revm{symbolic}{structured} and neural techniques.

\subsection{Trace Abstraction and Stateful Modeling}

Execution traces are notoriously large and semantically dense, making raw analysis costly and inaccessible. Prior work has proposed several abstraction techniques to reduce trace size while preserving interpretability. Pirzadeh et al.~\cite{pirzadeh2012trace} segment large traces into execution phases based on Gestalt principles and apply stratified sampling to enable scalable exploration. Feng et al.~\cite{feng2018hierarchical} introduce \emph{Sage}, a hierarchical abstraction framework that captures high-level behaviors from recurring low-level patterns. Hamou-Lhadj and Lethbridge propose summarization methods based on cue phrases and frequency analysis~\cite{hamou2005concept}, as well as the \emph{Utilityhood} metric~\cite{hamou2006summarizing} to rank components by structural relevance. Cornelissen et al.~\cite{cornelissen2008large} conduct a quantitative evaluation of multiple trace reduction techniques, such as sampling, stack depth limitation, and grouping. They examine the trade-offs of these methods in preserving behavioral fidelity.

Complementing these methods, the Trace Compass platform~\cite{kabamba2023node, hong2021trace, tracecompass_generic_state_system} provides a stateful abstraction via its \emph{State System}, which maintains a time-indexed database of attribute-value intervals for efficient querying. This system constructs a persistent model of trace state using event replays and interval trees, enabling scalable extraction of system-level behaviors~\cite{wininger2017declarative, zhang2023trace}. While these techniques offer powerful indexing and summarization, they lack semantic structuring and are limited in supporting flexible, user-driven queries over multidimensional trace entities.

\subsection{Knowledge Graphs for Execution Context Modeling}

Knowledge graphs (KGs) offer a semantically rich abstraction to represent system entities and their relationships. In trace contexts, they enable modeling of complex interactions among processes, threads, files, and CPUs in a form that is interpretable and machine-readable. Liang et al.~\cite{liang2024survey} categorize KGs into static, temporal, and multimodal types. Static KGs represent triples $(e_1, r, e_2)$ without time evolution; temporal KGs introduce evolving snapshots; and multimodal variants incorporate metrics or other data types such as images and videos. KG-based reasoning may be transductive, constrained to known entities, or inductive, supporting generalization. While KGs have been applied in domains like cybersecurity and software architecture, few systems construct them dynamically from execution traces. TAAF fills this gap by deriving time-indexed, per-query KGs from stateful trace outputs.

\subsection{LLMs for Structured Software Reasoning}

Large Language Models (LLMs) such as GPT-3~\cite{brown2020language} and GPT-4~\cite{achiam2023gpt} have demonstrated strong capabilities in NLP tasks like translation~\cite{lewis2019bart}, summarization~\cite{li2024pre}, and question answering~\cite{su2019generalizing}. However, their application to trace analysis faces two major barriers. First, trace data volumes often exceed the token limits of LLMs, making end-to-end ingestion infeasible. Second, LLMs encode knowledge implicitly and opaquely, leading to difficulties in validating results or understanding reasoning steps~\cite{pan2024unifying}. This often manifests as \emph{hallucinations} which refers to plausible-sounding but incorrect answers~\cite{ji2023survey}. Moreover, LLMs struggle with structured inputs such as graphs or tabular state histories, limiting their utility in domains requiring explicit causal reasoning.

\subsection{LLM–KG Integration Strategies}

To address LLM limitations, recent work explores hybrid approaches that integrate KGs into the language model pipeline. One dominant strategy is \textit{retrieval-augmented generation (RAG)}, which grounds LLMs in factually accurate subgraphs retrieved at inference time~\cite{baek2023kaping, feng2023knowledge}. Another line of work embeds KG structure directly into model architectures using entity-aware embeddings~\cite{zhang2019ernie, yamada2020luke} or graph neural modules such as QA-GNN~\cite{yasunaga2022qagnn}. A third family of techniques focuses on parameter-efficient domain adaptation using adapters~\cite{wang2020kadapter}, low-rank updates~\cite{hu2021lora}, or instruction-tuned prompts~\cite{sen2023knowledge}. These integrations improve factual accuracy and interpretability, but introduce latency and alignment challenges especially when operating on large, temporal traces~\cite{lewis2020retrieval, shi2023replug}. Recent efforts such as compressed history trees~\cite{kabamba2023node} aim to reduce memory and runtime costs of retrieval, motivating approaches like TAAF that combine high-performance indexing with \revm{symbolic}{} structuring and LLM reasoning.

\section{Methodology}
\label{sec:methodlogy}

\subsection{Problem Setting and Motivation}
\label{sec:problem-setting}
Modern kernel execution traces log millions of timestamped events across diverse entities such as threads, processes, CPUs, files, and network interfaces, offering rich runtime insights. However, answering practical trace-analysis questions from such voluminous and low-level data presents three core challenges. First, analysts must reason across varying temporal granularities, from instantaneous states to fine-grained intervals such as 1-second windows and full-trace trends. Second, many queries involve multidimensional interactions, for instance, reasoning across scheduling behavior, I/O flows, and resource contention, which requires interpretation of complex inter-entity relationships. Third, trace data lacks high-level semantic annotations, while analyst queries range from simple factual lookups to comparative or causal questions, demanding both flexible and precise semantic understanding. We also wanted to see how far we could go in trace analysis with LLMs and knowledge graphs.

\begin{figure*}[t]
  \centering
  \includegraphics[width=.85\textwidth]{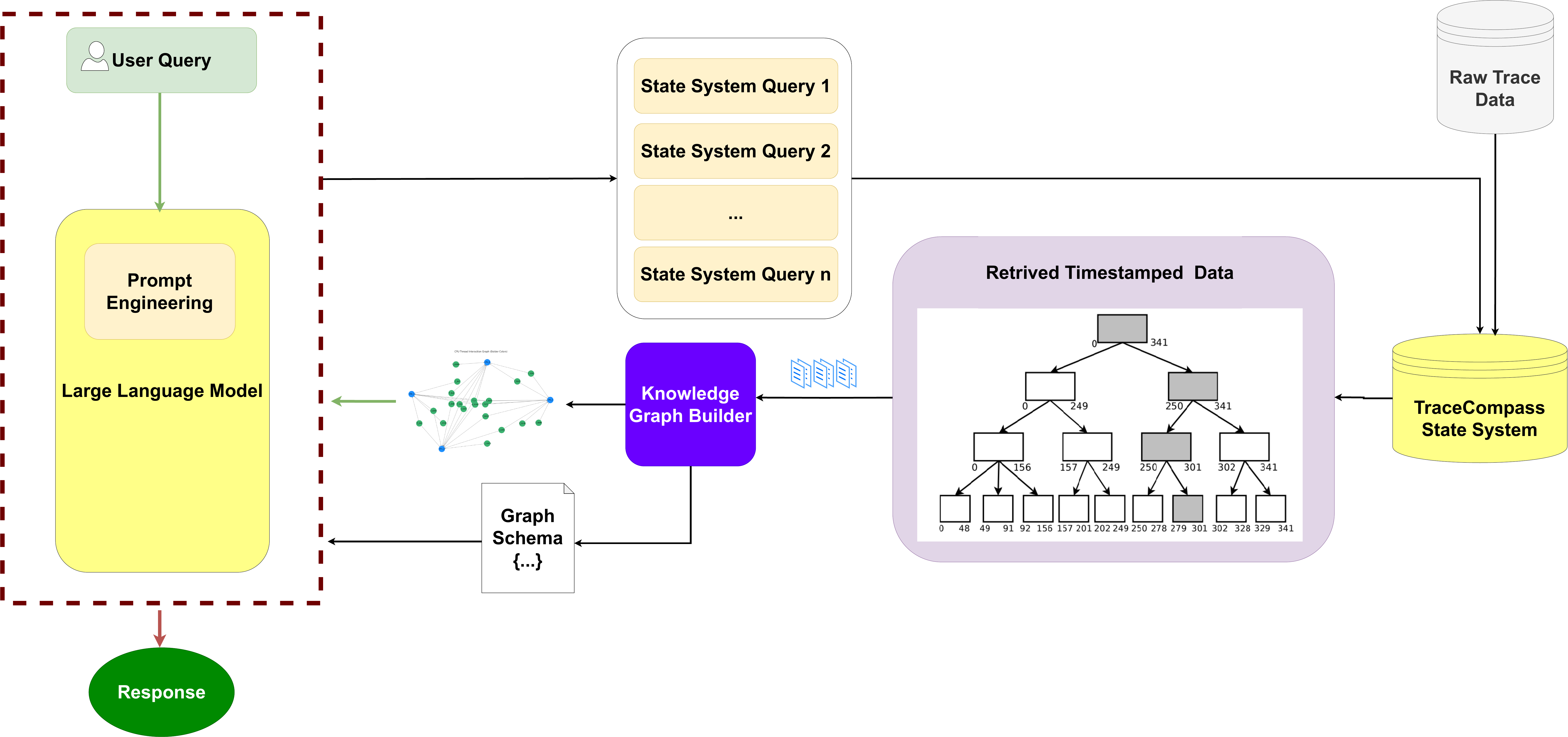}
  \caption{End-to-end architecture \rev{overview} of the Trace Abstraction and Analysis Framework (TAAF).}
  \label{fig:taaf}
\end{figure*}

To address these challenges, we introduce the Trace Abstraction and Analysis with Foundation Models (TAAF) framework, which integrates scalable data abstraction, structured knowledge representation, and natural-language reasoning to enable flexible, accurate, and explainable trace analysis. \rev{Figure~\ref{fig:taaf} illustrates the framework architecture overview.}
\subsection{Architectural Overview and Trace-to-Answer Pipeline}
\label{sec:taaf-overview}

TAAF is a three-layered abstraction framework for \revm{symbolic}{structured} trace analysis. Rather than analyzing raw kernel trace events directly, TAAF systematically transforms them into structured, interpretable representations suitable for \revm{symbolic}{structured} reasoning and LLM-based answering.

Overall, the system defines a trace-to-answer transformation pipeline $\mathcal{T} \rightarrow \mathcal{S} \rightarrow \mathcal{G} \rightarrow \mathcal{A}$, where $\mathcal{T}$ is the raw trace, $\mathcal{S}$ is the State System, $\mathcal{G}$ is the knowledge graph, and $\mathcal{A}$ is the final answer. Each layer serves to constrain scope, preserve structure, and expose trace semantics to the model in a usable form.


\begin{itemize}
    \item \textbf{Layer 1: From Trace to State.} The raw event stream $\mathcal{T}$ is transformed into a \revm{symbolic}{structured}, time-indexed State System $\mathcal{S} = (Q, H)$ using a deterministic encoding function:
    \[
    \mathcal{T} \xrightarrow{\phi_1} \mathcal{S}
    \]
    This transformation enables scalable temporal reasoning by organizing system attributes and their state transitions as synchronized trees.

    \item \textbf{Layer 2: From State to Graph.} Given a natural-language query $q$, TAAF extracts relevant temporal intervals from $\mathcal{S}$ and constructs a semantic knowledge graph $\mathcal{G}_q$ that represents system interactions within the scope of the query:
    \[
    \mathcal{S}, q \xrightarrow{\phi_2} \mathcal{G}_q
    \]
    This graph is query-specific, minimal, and enriched with temporal and semantic metadata.

    \item \textbf{Layer 3: From Graph to Answer.} The graph $\mathcal{G}_q$ and an associated schema prompt $\mathcal{C}_q$ are serialized and passed to a large language model, which produces a grounded natural-language answer $\mathcal{A}_q$:
    \[
    \mathcal{G}_q, \mathcal{C}_q, q \xrightarrow{\phi_3} \mathcal{A}_q
    \]
    This LLM-based reasoning phase supports quantitative analysis, comparisons, and explanations grounded in trace semantics.
\end{itemize}


\revm{We refer to these intermediate representations as symbolic because they capture system behavior in structured, semantically annotated forms, supporting reasoning, composability, and extensibility across trace types.} {These intermediate representations are structured and semantically annotated; they capture system behavior and support reasoning, composability, and extensibility across trace types.}

We revisit this layered design in our evaluation (Section~\ref{sec:analysis}) to assess its impact on scalability, semantic grounding, and answer quality.

We describe each layer in more details below.
\subsection{Layer 1: From Raw Trace to State System Abstraction}
\label{sec:layer1}

A kernel trace $\mathcal{T}$ is defined as a temporally ordered set of fine-grained system events:
\[
\mathcal{T} = \{ (e_i, t_i, c_i) \mid e_i \in \mathcal{E},\ t_i \in \mathbb{T},\ c_i \in \mathcal{C} \}
\]
Each tuple consists of an event type $e_i$, a timestamp $t_i$, and a set of involved system components $c_i$ such as threads, CPUs, locks, or I/O devices. These events reflect low-level operating system behavior, including context switches, file operations, and synchronization events. While highly expressive, such traces are both voluminous and semantically sparse. Even a few seconds of system activity may yield tens of millions of events, rendering direct reasoning infeasible due to the combined temporal complexity, multidimensionality of interacting entities, and lack of structured semantics.

To address this, we define a formal transformation $\phi_1$ that converts a raw trace $\mathcal{T}$ into a time-indexed \revm{symbolic}{structured} abstraction known as the \textit{State System}. Originally proposed by Montplaisir et al.~\cite{montplaisir2013efficient}, the State System enables scalable querying and semantic enrichment by organizing the trace into two synchronized tree structures: an \emph{attribute tree} and a \emph{history tree}.

Formally, the State System is defined as $\mathcal{S} = (Q, H)$, where:
\begin{itemize}
    \item $Q = \{q_1, q_2, \ldots, q_n\}$ is the set of stable integer identifiers (\emph{quarks}) corresponding to hierarchical attribute paths (e.g., \texttt{/CPU/2/Thread/5130}),
    \item $H$ is the history tree, recording value transitions as:
    \[
    H = \left\{ \langle t_{\text{start}}, t_{\text{end}}, q, v \rangle \mid q \in Q, v \in \mathcal{V} \right\}
    \]
    Each entry indicates that attribute $q$ held value $v$ during the interval $[t_{\text{start}}, t_{\text{end}})$.
\end{itemize}

The transformation from raw events to this structure is governed by a predefined set of event-type-specific functions. For each event type $e \in \mathcal{E}$, we define a mapping:
\[
\phi_1^{(e)} : \mathbb{T} \times \mathcal{C} \rightarrow \mathcal{P}(H)
\]
which encodes how an event of type $e$, given its timestamp and involved components, contributes to zero or more entries in $H$. The complete transformation $\phi_1$ is the union of these specialized rules:
\[
\phi_1(\mathcal{T}) = \bigcup_{(e_i, t_i, c_i) \in \mathcal{T}} \phi_1^{(e_i)}(t_i, c_i)
\]

This formulation captures the deterministic nature of event-to-state updates while preserving modularity across heterogeneous event types.

This dual-tree design supports efficient \revm{symbolic}{structured} queries such as:
\[
\text{query}(q, t) \rightarrow v \quad \text{and} \quad \text{query}(q, [t_1, t_2]) \rightarrow \{v_i\}
\]
enabling localized reasoning over fine-grained execution intervals without requiring full trace scans.

By compactly encoding temporal dynamics in this form, the State System allows higher-level semantic queries to be executed in logarithmic time relative to the trace size~\cite{montplaisir2013efficient,kabamba2023node,zhang2023trace}. This layer thus transforms $\mathcal{T}$ into a \revm{symbolic,}{}semantically enriched structure $\mathcal{S}$ that forms the foundation for downstream reasoning tasks in TAAF.

\subsection{Layer 2: From State System to Query-Specific Knowledge Graph}
\label{sec:layer2}

Given the \revm{symbolic}{structured} state representation $\mathcal{S} = (Q, H)$ produced by $\phi_1$, the second layer of TAAF constructs a dynamic, query-specific knowledge graph that grounds trace semantics in a structured and interpretable form. This transformation is handled by a function $\phi_2$, which takes the State System $\mathcal{S}$ and a natural language query $q$ as input, and produces a query-specific knowledge graph $\mathcal{G}_q$.

\begin{figure}[t]
  \centering
  \includegraphics[width=.8\columnwidth]{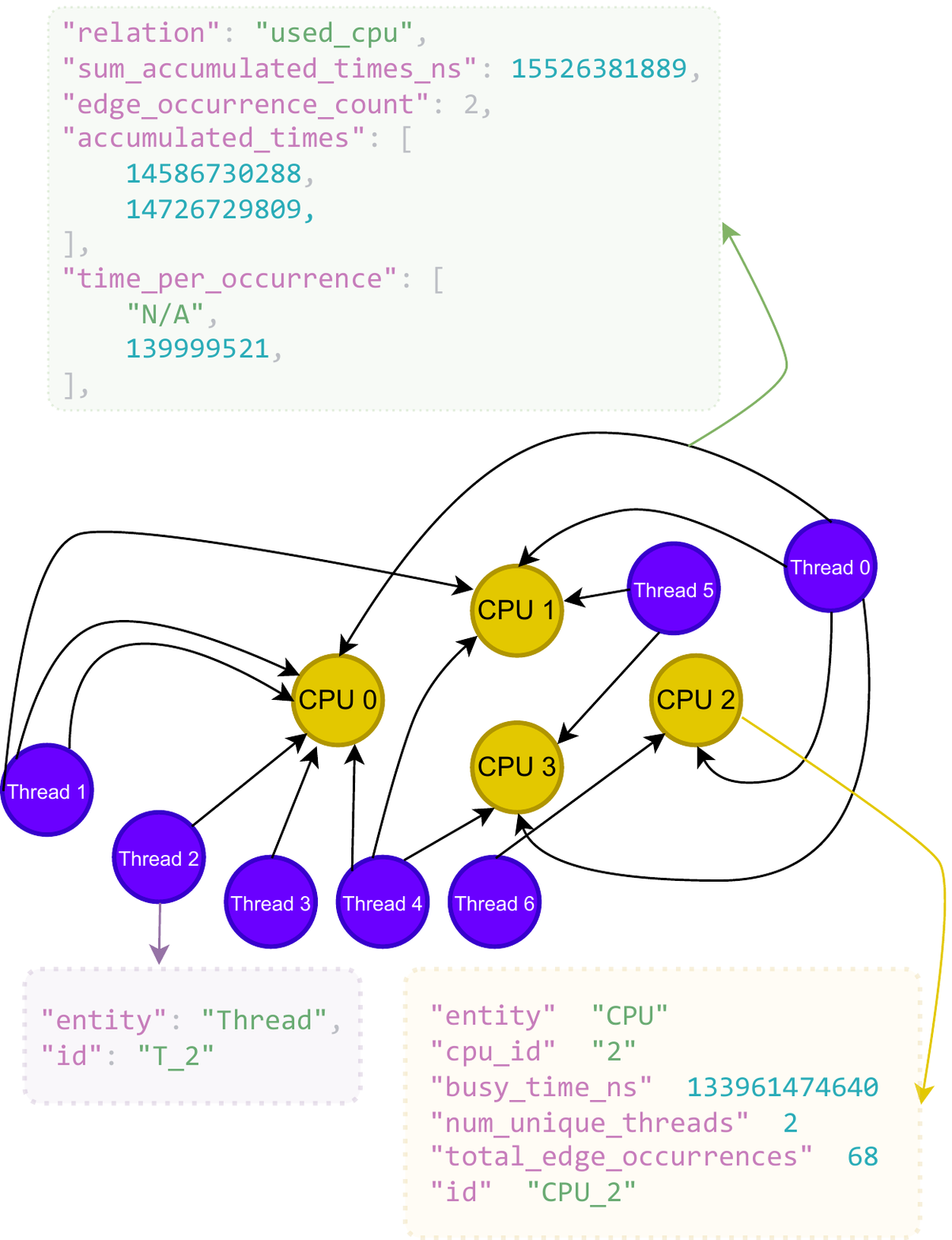}
  \caption{Example of a query-specific knowledge graph generated from a 1-second trace interval.
  }
  \label{fig:kg-example}
\end{figure}

Formally, we define $\mathcal{G}_q$ a 4-tuple:
\[
\mathcal{G}_q = (V_q, E_q, \mu_q, \tau_q)
\]
where $V_q$ denotes the set of typed nodes representing system entities, and $E_q \subseteq V_q \times \mathcal{L} \times V_q$ denotes the set of labeled, directed edges that encode semantic relationships between those entities. The label space $\mathcal{L}$ includes relation types such as \texttt{executes\_on}, \texttt{reads\_from}, and \texttt{holds\_lock}. Each edge $e \in E_q$ is assigned a weight via the function $\mu_q : E_q \rightarrow \mathbb{R}$, typically representing quantitative values such as duration or frequency. Temporal scope is encoded through a mapping $\tau_q: V_q \cup E_q \rightarrow \mathbb{T}$, which associates each entity or interaction with a time window derived from $H$.

This knowledge graph is constructed on demand, scoped narrowly to the query $q$ and the relevant interval within the trace. For example, if $q$ asks, “How long did Thread 5130 run on CPU 2 between $t_1$ and $t_2$?”, the system identifies the relevant quarks from $Q$, queries their associated intervals in $H$, and maps the result to a graph with two nodes (representing Thread 5130 and CPU 2) and a single edge labeled \texttt{executes\_on}, weighted by total runtime (e.g., $\mu(e) = 5.25$ seconds) and scoped to the interval $\tau(e) = [t_1, t_2]$. Figure~\ref{fig:kg-example} shows an example KG generated from question querying a 1-second trace window.

By extracting only the entities and relations needed for $q$, $\mathcal{G}_q$ avoids the overgeneralization and interpretability issues of global or static graphs. It supports \revm{symbolic}{structured} reasoning in a lightweight, interpretable, and semantically aligned form. Edges are directed to capture causal relationships and weighted with quantitative metrics from the State System. The graph is tailored to each query and expandable with additional metadata to enable focused, quantitative, and adaptable analysis grounded in domain knowledge.


This step completes the semantic grounding layer of TAAF, translating time-series trace data into a \revm{symbolic}{structured} graph format suitable for large language model reasoning.

\subsection{Layer 3: From Knowledge Graph to Answer via LLMs}
\label{sec:layer3}

The final layer of TAAF performs semantic reasoning over the \revm{symbolic}{structured} representation $\mathcal{G}_q$ to generate a natural-language answer $\mathcal{A}_q$. This transformation is governed by the function $\phi_3$, which takes as input the knowledge graph $\mathcal{G}_q$, a natural-language query $q$, and a schema prompt $\mathcal{C}_q$ that explicitly encodes the meaning of graph elements. The output is a grounded, human-readable answer:
\[
\mathcal{A}_q = \phi_3(\mathcal{G}_q, \mathcal{C}_q, q)
\]

To maximize the factual accuracy and interpretability of LLM responses, TAAF follows a structured prompting strategy. The schema prompt $\mathcal{C}_q$ formally defines the types of nodes and edges in $\mathcal{G}_q$, their attributes (e.g., execution time, file path), and any additional domain-specific hints. The graph $\mathcal{G}_q$ itself is serialized in a structured JSON format, preserving the graph topology and annotated metadata. Together, the components are passed as input to the large language model alongside the query $q$ in a templated format:

\begin{quote}
\ttfamily
\{
  "schema": \{...\},\\
  "graph": \{...\},\\
  "\textbf{user query}": "Which thread executed most on CPU 1 between $t_1$ and $t_2$?"\\
\}
\end{quote}

The model is thus explicitly grounded in a well‑scoped semantic context, minimizing the need for open‑ended inference. The result is the answer $\mathcal{A}_q$, which may include the direct outcome. Moreover, it can contain justifications or comparative insights, depending on the query type and model capabilities.

For example, given the above user query, the model may return:

\begin{quote}
\emph{“Thread 9127 executed for 6.4 seconds on CPU 1 between $t_1$ and $t_2$, which is the highest among all threads during that interval.”}
\end{quote}

This output demonstrates the core strength of the TAAF framework. By separating trace understanding into a \revm{symbolic}{structured} abstraction layer and deferring reasoning to an LLM that operates over this structured input, we support robust, explainable trace analysis without relying on static rules or brittle heuristics.

Importantly, the LLM interface in TAAF is designed to be model-agnostic. In our implementation, we employ both general-purpose and reasoning-tuned foundation models (e.g., \texttt{o4-mini}), and observe that schema-conditioned inputs significantly improve factual precision. Furthermore, the prompt construction process is transparent and auditable, making it easier to debug reasoning failures or extend the system to new query types or domains.

By combining \revm{symbolic}{the} structure with foundation model reasoning, this final layer completes the trace-to-answer pipeline and enables TAAF to flexibly support diverse analytic needs across performance diagnosis, behavioral explanation, and root cause investigation.

\section{Empirical Evaluation}
\label{sec:analysis}
This section details the experimental design, benchmark construction, evaluation protocol, scoring metrics, research questions, and the results that collectively assess the accuracy, robustness, and explainability of TAAF.

\subsection{Evaluation Framework: \textsc{TraceQA-100}}
\label{sec:benchmark}

Despite progress in using AI and LLM for trace analysis, the community lacks a public, ground-truth dataset for reasoning over kernel-level execution traces. Existing LLM benchmarks such as MMLU or Big-Bench do not address the unique challenges of trace analysis, including fine-grained time, multi-entity interactions, and numeric aggregation. To fill this gap, we introduce \textsc{TraceQA--100} 
 a curated benchmark designed to exercise all three dimensions.

\textbf{Trace provenance and slicing.}  
We use LTTng traces of the SciMark 2.0 Java benchmark~\cite{scimark2020}, which generates approximately 34\,M kernel events. From each run, we extract 1s, 10s, and 100s slices around three canonical temporal locations, \emph{start}, \emph{mid}, and \emph{end}, yielding a pool of trace segments that serve as the factual basis for question construction.

\textbf{Question authorship.}  
Questions were generated through a four-step, double-blind process. Two experts independently inspected trace slices and drafted questions requiring time-aware, multi-entity reasoning. Drafts were peer-reviewed and normalized into three answer formats: explanatory, multiple-choice, or true/false. A third expert, blind to the drafts, produced reference answers using hand-written Trace Compass scripts; disagreements were resolved through adjudication. Each question was then tagged as \emph{single-\rev{hop}} or \emph{multi-\rev{hop}} based on the reasoning scope.

The resulting set includes 100 questions, evenly split into 40 explanatory, 30 multiple-choice, and 30 true/false items, and a 50/50 split across \rev{hop} scopes. Since each prompt is instantiated on nine distinct trace slices (3 temporal locations × 3 time-window lengths), the benchmark comprises $100\times9=900$ unique question–trace-segment pairs. Representative examples are shown in Table~\ref{tab:traceqa_sample}.

\textbf{Schema and reproducibility.}  
Each item in \textsc{TraceQA-100} includes the textual prompt, trace identifier, answer format, reference answer, and relevant metadata. All answer scripts are shipped and re-executed as part of our CI pipeline to verify trace-consistency. Detailed labeling guidelines are included to support extensions beyond SciMark or Linux.

To our knowledge, \textsc{TraceQA-100} is the first public benchmark that pairs large-scale kernel traces with expert-verified QA tasks. By targeting temporal, structural, and numeric reasoning, it offers a reusable reference point for future \revm{symbolic}{structured} or neural approaches to trace understanding.

\begin{table*}[t]
\centering
\small
\begin{tabular}{llllll}
\toprule
\textbf{Question type} & \textbf{\rev{hop} scope} & \textbf{Example question} &
\textbf{Temporal loc} & \textbf{Time window} & \textbf{Ref answer} \\
\midrule
Explanation & Single & What is the total accumulated CPU time for \texttt{thread 5130} on CPU 0? & mid & 1 s & 5,247,200 ns \\
Explanation & Multi  & Which CPU served the most number of distinct threads? & mid & 100 s & CPU\_0 \\
Multiple choice & Single & Thread 2559 primarily uses: (A) CPU\_0 (B) CPU\_1 (C) CPU\_2 (D) None & start & 10 s & B \\
Multiple choice & Multi  & Thread 5236 primarily uses: (A) CPU\_0 (B) CPU\_1 (C) CPU\_2 (D) CPU\_3 & end & 10 s & C \\
True or False & Single & True or False? CPU 2 total busy time > 1e11 ns & mid & 10 s & False \\
True or False & Multi  & True or False? The busiest CPU also runs the most unique threads & start & 10 s & True \\
\bottomrule
\end{tabular}
\caption{Sample items from TraceQA-100 covering question formats, \rev{hop} scopes, temporal locations, and windows.}
\label{tab:traceqa_sample}
\end{table*}

\begin{table*}[t]
  \centering
  \small

  \label{tab:eval_configs}
  \begin{tabular}{@{}p{2.8cm}p{5.4cm}p{8.8cm}@{}}
    \toprule
    \textbf{Variant} & \textbf{Input passed to LLM} & \textbf{Practical outcome} \\ \midrule
    Events-only & Raw kernel events ($\ge\!10^{7}$) & Exceeds context limits; produces incoherent answers.  Retained only to illustrate the need for abstraction. \\
    Baseline & State-System numeric values & Fits budget but remains low-level. \\
    TAAF & Query-specific temporal knowledge graph & Adds explicit entities and relations; grounds the model while staying compact. \\ \bottomrule
  \end{tabular}
    \caption{Trace representations supplied to the language model.  Only the latter two fit within contemporary context windows.}
    \label{tab:eval_configs}
\end{table*}

\subsection{Evaluation Setup and Metrics}
\label{sec:evaluation-setup}

Each item in \textsc{TraceQA-100} is a standalone QA task, combining a natural-language prompt, a temporal slice of the SciMark\,2.0 trace, and a single ground-truth answer. We evaluate each task under three trace–representation settings (Table~\ref{tab:eval_configs}): the \emph{events-only} variant streams raw kernel events and serves as a qualitative baseline; the \emph{Baseline} uses the numeric output of the temporal index; and \textbf{TAAF} augments this with a query-specific knowledge graph built by $\phi_2$.

To account for LLM stochasticity, we follow best practices~\cite{lior2025reliableeval} and sample each (question, representation, model) combination three times. With 100 benchmark questions, this yields $300$ scored responses per configuration. Across three models, three time windows (1\,s, 10\,s, 100\,s), and two structured inputs (Baseline, TAAF), the core grid totals $5{,}400$ labeled outputs. It should be noted that events-only runs are included only for qualitative comparison, as such runs typically exceed the LLM context window and thus are not part of the scalable evaluation. \\

All runs execute on a single GPU node. End-to-end processing of the largest 100-second trace, including indexing, KG construction, and inference, completes in under 40 seconds, confirming feasibility for interactive use.

Each response is labeled on a 3-point scale: a score of \textbf{0} indicates the answer is incorrect or irrelevant; \textbf{0.5} denotes a partially correct response with minor errors or incomplete reasoning; and \textbf{1} signifies a fully correct answer that is consistent with the ground truth.

We report two metrics. \textbf{Accuracy} is the weighted average over scores:
\begin{equation}
\text{Accuracy} = \frac{(N_0 \cdot 0) + (N_{0.5} \cdot 0.5) + (N_1 \cdot 1)}{N_{\text{total}}} \times 100
\end{equation}

\noindent where \(N_0, N_{0.5}, N_1\) denote the count of each label (out of \(N_{\text{total}} = 300\)).

\textbf{Consistency} captures the model's stability over repeated samples. We compute entropy:
\begin{equation}
E = -\sum_{i \in \{0, 0.5, 1\}} P_i \log_2 P_i
\end{equation}
\begin{equation}
\text{Consistency} = \left(1 - \frac{E}{\log_2 3}\right) \times 100
\end{equation}

\noindent where \(P_i\) is the empirical frequency of score \(i\). Higher consistency indicates lower variance across trials.

\paragraph{Model Configuration.}
We evaluate three models via API: \textbf{GPT-4.1 nano}, a compact high-throughput model; \textbf{GPT-4o}, OpenAI’s current flagship; and \textbf{o4-mini (Reasoning)}, optimized for high reasoning accuracy. All models use the same prompt template and few-shot examples. Decoding temperature is fixed at 0.5, except in RQ7. For o4-mini, we activate \texttt{reasoning=high}.

\rev{\textit{Cross-family probe.} To address generalizability beyond a single vendor, we also evaluate Google and Anthropic models. From Google, we use \textbf{Gemini 2.5 Pro} as the flagship reasoning model and \textbf{Gemini 2.5 Flash} as the fast, cost-efficient workhorse. From Anthropic, we use \textbf{Claude Opus 4.1} as the flagship reasoning model and \textbf{Claude Haiku 4.5} as the fast, cost-efficient model. We selected these pairs because each represents its vendor’s top reasoning model alongside a smaller, speed- and value-oriented model, and both vendors publicly position them as such. All four models use the same prompt template and scoring protocol as the GPT models.}

\subsection{Research Questions}

To guide our evaluation, we formulate seven research questions grouped into two phases. Phase 1 explores core performance dimensions of TAAF across query types, time windows, and LLM variants. Phase 2 isolates specific factors such as schema inclusion, temporal placement, and sampling behavior.

\textbf{Phase 1 – Full-grid evaluation}
\begin{itemize}[leftmargin=1em]
    \item \textbf{RQ1:} How does TAAF's accuracy vary across different user query types, such as multiple-choice, true/false, and explanatory, and across graph question types, such as single-\rev{hop} versus multi-\rev{hop}?
    \item \textbf{RQ2:} What is the effect of time-window length on the quality of model responses?
    \item \textbf{RQ3:} To what extent does incorporating a knowledge graph improve accuracy compared to raw State System output?
    \item \textbf{RQ4:} How does performance vary across LLM backends \rev{including a focused cross-family probe}?
\end{itemize}
\textbf{Phase~1} spans a 3×3 grid of evaluation settings formed by three LLMs and three time-window lengths (1s, 10s, 100s), all sampled from the midpoint of the trace. For each cell in the grid, we run both the TAAF pipeline with its query-specific knowledge graph and the baseline pipeline without graph grounding. Each configuration covers 100 benchmark questions sampled three times, yielding $9 \times 2 \times 300 = 5400$ labeled responses that inform RQ1 through RQ4.

\textbf{Phase 2 – Focused factors}
\begin{itemize}[leftmargin=1em]
    \item \textbf{RQ5:} Does providing the LLM with the graph schema, such as node types and features, enhance the accuracy of its responses?
    \item \textbf{RQ6:} To what extent does the temporal location of the window (early, mid, late) affect accuracy?
    \item \textbf{RQ7:} How does the temperature parameter(sampling randomness) influence both accuracy and response consistency?
\end{itemize}

\textbf{Phase~2} holds the model (GPT-4o) and query grounding fixed, then varies key factors. RQ5 toggles schema support in the mid-trace 1s slice, contributing 300 responses. RQ6 compares temporal window placement, early versus late in the trace, at the 10s granularity, adding 600 responses. RQ7 examines the impact of sampling temperature by re-running the mid-trace 10s slice at four values (0.1, 0.3, 0.7, 0.9) for an additional 1\rev{,}200 responses. In total, Phase~2 adds 2\rev{,}100 outputs, bringing the complete dataset to 7\rev{,}500 labeled responses used in the following analysis.

The empirical results for these research questions are presented in the following.

\subsection{Phase 1: Full Grid Experiments}
\textbf{RQ1: How does TAAF's accuracy vary across different user query types, and across graph question types, such as single-\rev{hop} versus multi-\rev{hop}?}

We begin the analysis by probing how user and graph question styles influence TAAF’s success. Our TraceQA-100 benchmark mixes three natural-language formats: (1) True/False, (2) Multiple Choice, and (3) Explanatory.  
Each format is asked in two graph-question settings: (1) single-\rev{hop} questions that focus on one central entity, such as a CPU, and its local neighbourhood; (2) multi-\rev{hop} questions that require reasoning about multiple distinct centres, for example two CPUs that share or migrate threads.
\begin{figure*}[t]
  \centering
  \includegraphics[width=\textwidth]{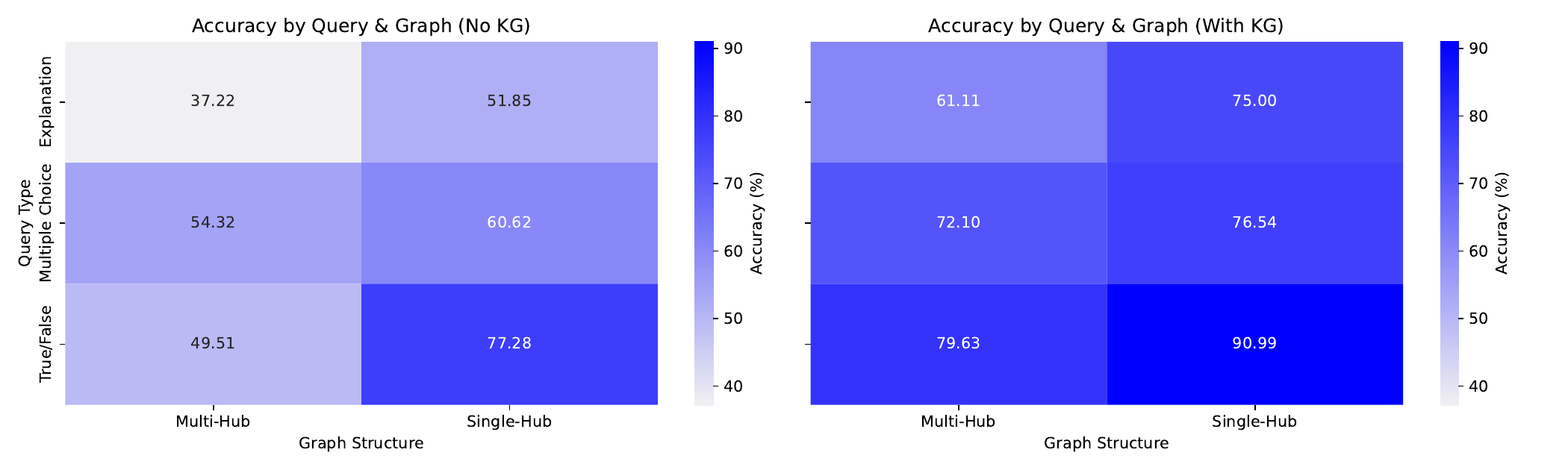}
  \caption{Accuracy by query type and \rev{hop} count. Left: baseline; right: TAAF. White text shows accuracy gain (+p.p.).}
  \label{fig:rq4_heatmap}
\end{figure*}

\begin{table}[ht]
  \centering
  \scriptsize
  \caption{Accuracy  results for LLM models across different intervals and query types, with and without a KG.}
  \label{tab:combined_results}
  \footnotetext{Abbreviations: UQT = User Query Type; GQT = Graph Question Type; Acc\% = weighted accuracy; S-\rev{hop} = single-\rev{hop} question; M-\rev{hop} = multi-\rev{hop} question; MC = multiple choice; T/F = true–false; Expl. = explanatory.}
  \resizebox{\columnwidth}{!}{%
    \rowcolors{3}{gray!15}{white}
    \begin{tabular}{llllcccccccc}
      \toprule
      \rowcolor{gray!30}
      \textbf{Model} & \textbf{Interval} & \textbf{UQT} & \textbf{GQT} &
      \multicolumn{4}{c}{\textbf{No KG (Baseline)}} &
      \multicolumn{4}{c}{\textbf{With KG (TAAF)}} \\
      \cmidrule(lr){5-8}\cmidrule(lr){9-12}
      & & & & 0\% & 0.5\% & 1\% & Acc\% & 0\% & 0.5\% & 1\% & Acc\% \\
      \midrule
      GPT 4.1 nano & 1 s   & Expl. & S-\rev{hop} & 45.00 & 13.33 & 41.67 & 48.33 & 33.33 &  8.33 & 58.33 & 62.50 \\
      GPT 4.1 nano & 1 s   & Expl. & M-\rev{hop} & 61.67 & 20.00 & 18.33 & 28.33 & 46.67 & 20.00 & 33.33 & 43.33 \\
      GPT 4.1 nano & 1 s   & MC    & S-\rev{hop} & 35.56 & 17.78 & 46.67 & 55.56 & 11.11 &  5.56 & 83.33 & \textbf{86.11} \\
      GPT 4.1 nano & 1 s   & MC    & M-\rev{hop} & 40.00 & 11.11 & 48.89 & 54.44 & 26.67 &  4.44 & 68.89 & 71.11 \\
      GPT 4.1 nano & 1 s   & T/F   & S-\rev{hop} & 20.00 &  8.89 & 71.11 & 75.56 & 10.00 &  2.22 & 87.78 & \textbf{88.89} \\
      GPT 4.1 nano & 1 s   & T/F   & M-\rev{hop} & 41.11 & 21.11 & 37.78 & 48.33 & 16.67 & 12.22 & 71.11 & 77.22 \\
      \cmidrule(lr){2-12}
      GPT 4.1 nano & 10 s  & Expl. & S-\rev{hop} & 46.67 & 16.67 & 36.67 & 45.00 & 43.33 &  5.00 & 51.67 & 54.17 \\
      GPT 4.1 nano & 10 s  & Expl. & M-\rev{hop} & 56.67 & 21.67 & 21.67 & 32.50 & 46.67 & 13.33 & 40.00 & 46.67 \\
      GPT 4.1 nano & 10 s  & MC    & S-\rev{hop} & 50.00 & 13.33 & 36.67 & 43.33 & 29.17 &  5.83 & 65.00 & 67.92 \\
      GPT 4.1 nano & 10 s  & MC    & M-\rev{hop} & 55.67 & 13.00 & 31.33 & 37.83 & 37.14 &  4.29 & 58.57 & 60.71 \\
      GPT 4.1 nano & 10 s  & T/F   & S-\rev{hop} & 27.78 &  9.44 & 62.78 & 67.50 & 11.11 &  4.44 & 84.44 & 86.67 \\
      GPT 4.1 nano & 10 s  & T/F   & M-\rev{hop} & 55.56 & 16.67 & 27.78 & 36.11 & 29.17 &  9.17 & 61.67 & 66.25 \\
      \cmidrule(lr){2-12}
      GPT 4.1 nano & 100 s & Expl. & S-\rev{hop} & 49.03 & 15.97 & 34.99 & 42.98 & 45.10 &  7.84 & 47.06 & 50.00 \\
      GPT 4.1 nano & 100 s & Expl. & M-\rev{hop} & 59.15 & 21.13 & 19.72 & 30.29 & 49.30 & 19.72 & 30.99 & 40.85 \\
      GPT 4.1 nano & 100 s & MC    & S-\rev{hop} & 43.40 & 19.81 & 36.79 & 46.70 & 26.61 &  6.42 & 66.97 & 70.18 \\
      GPT 4.1 nano & 100 s & MC    & M-\rev{hop} & 48.50 & 13.50 & 38.00 & 44.75 & 33.58 &  6.72 & 59.70 & 62.06 \\
      GPT 4.1 nano & 100 s & T/F   & S-\rev{hop} & 35.63 & 14.38 & 50.00 & 57.19 & 14.13 &  6.38 & 79.50 & \textbf{82.69} \\
      GPT 4.1 nano & 100 s & T/F   & M-\rev{hop} & 55.10 & 17.65 & 27.25 & 36.07 & 26.13 & 10.92 & 62.95 & 67.41 \\
      \midrule
      GPT 4o        & 1 s   & Expl. & S-\rev{hop} & 30.00 &  8.00 & 62.00 & 66.00 & 18.00 &  2.67 & 79.33 & \textbf{80.67} \\
      GPT 4o        & 1 s   & Expl. & M-\rev{hop} & 46.00 & 12.00 & 42.00 & 48.00 & 29.33 &  9.33 & 61.33 & 66.00 \\
      GPT 4o        & 1 s   & MC    & S-\rev{hop} & 26.00 &  6.00 & 68.00 & 71.00 &  7.00 &  1.33 & 91.67 & \textbf{92.33} \\
      GPT 4o        & 1 s   & MC    & M-\rev{hop} & 33.33 &  8.33 & 58.33 & 62.50 & 13.33 &  3.33 & 83.33 & \textbf{85.00} \\
      GPT 4o        & 1 s   & T/F   & S-\rev{hop} & 13.33 &  5.00 & 81.67 & \textbf{84.17} &  4.00 &  0.67 & 95.33 & \textbf{95.67} \\
      GPT 4o        & 1 s   & T/F   & M-\rev{hop} & 32.00 & 14.67 & 53.33 & 60.67 & 12.67 &  3.33 & 84.00 & \textbf{85.67} \\
      \cmidrule(lr){2-12}
      GPT 4o        & 10 s  & Expl. & S-\rev{hop} & 34.48 & 11.49 & 54.02 & 59.76 & 33.33 &  3.45 & 63.22 & 64.94 \\
      GPT 4o        & 10 s  & Expl. & M-\rev{hop} & 50.00 & 12.50 & 37.50 & 43.75 & 38.46 & 11.54 & 50.00 & 55.77 \\
      GPT 4o        & 10 s  & MC    & S-\rev{hop} & 34.55 & 12.73 & 52.73 & 58.09 & 13.64 &  2.73 & 83.64 & \textbf{85.00} \\
      GPT 4o        & 10 s  & MC    & M-\rev{hop} & 46.43 & 11.90 & 41.67 & 47.62 & 22.32 &  3.57 & 74.11 & 75.89 \\
      GPT 4o        & 10 s  & T/F   & S-\rev{hop} & 18.27 &  6.73 & 75.00 & 78.37 &  6.25 &  2.50 & 91.25 & \textbf{92.50} \\
      GPT 4o        & 10 s  & T/F   & M-\rev{hop} & 41.67 & 16.67 & 41.67 & 50.00 & 17.44 & 10.47 & 72.09 & 76.32 \\
      \cmidrule(lr){2-12}
      GPT 4o        & 100 s & Expl. & S-\rev{hop} & 39.39 & 12.12 & 48.48 & 54.55 & 29.54 &  8.41 & 62.05 & 66.26 \\
      GPT 4o        & 100 s & Expl. & M-\rev{hop} & 55.17 & 15.52 & 29.31 & 36.07 & 35.79 & 14.74 & 49.47 & 56.84 \\
      GPT 4o        & 100 s & MC    & S-\rev{hop} & 39.22 & 14.71 & 46.08 & 53.44 & 11.76 &  3.92 & 84.31 & \textbf{86.27} \\
      GPT 4o        & 100 s & MC    & M-\rev{hop} & 50.88 & 14.04 & 35.09 & 42.11 & 23.81 &  5.95 & 70.24 & 72.22 \\
      GPT 4o        & 100 s & T/F   & S-\rev{hop} & 26.92 &  9.62 & 63.46 & 68.27 &  7.21 &  2.88 & 89.90 & \textbf{91.35} \\
      GPT 4o        & 100 s & T/F   & M-\rev{hop} & 47.83 & 17.39 & 34.78 & 43.48 & 15.85 &  6.71 & 77.44 & \textbf{80.80} \\
      \midrule
      GPT o4-mini   & 1 s   & Expl. & S-\rev{hop} & 17.78 & 15.56 & 66.67 & 74.44 &  4.44 &  5.00 & 90.56 & \textbf{93.06} \\
      GPT o4-mini   & 1 s   & Expl. & M-\rev{hop} & 35.24 & 22.86 & 41.90 & 53.33 &  8.57 & 14.29 & 77.14 & \textbf{84.29} \\
      GPT o4-mini   & 1 s   & MC    & S-\rev{hop} & 17.02 &  8.51 & 74.47 & 78.72 &  3.19 &  1.06 & 95.74 & \textbf{96.27} \\
      GPT o4-mini   & 1 s   & MC    & M-\rev{hop} & 21.21 & 14.14 & 64.65 & 71.72 &  5.41 &  2.70 & 91.89 & \textbf{93.24} \\
      GPT o4-mini   & 1 s   & T/F   & S-\rev{hop} &  8.33 &  7.22 & 84.44 & \textbf{88.06} &  2.22 &  1.11 & 96.67 & \textbf{97.22} \\
      GPT o4-mini   & 1 s   & T/F   & M-\rev{hop} & 21.75 & 13.75 & 64.50 & 71.88 &  4.37 &  2.46 & 93.17 & \textbf{94.35} \\
      \cmidrule(lr){2-12}
      GPT o4-mini   & 10 s  & Expl. & S-\rev{hop} &  9.52 & 13.10 & 77.38 & \textbf{83.93} &  5.95 &  5.95 & 88.10 & \textbf{91.07} \\
      GPT o4-mini   & 10 s  & Expl. & M-\rev{hop} & 27.40 & 19.86 & 52.74 & 62.67 &  9.59 & 13.70 & 76.71 & \textbf{83.56} \\
      GPT o4-mini   & 10 s  & MC    & S-\rev{hop} & 13.68 & 10.26 & 76.06 & \textbf{81.19} &  3.08 &  3.08 & 93.85 & \textbf{95.38} \\
      GPT o4-mini   & 10 s  & MC    & M-\rev{hop} & 18.10 & 11.43 & 70.48 & 76.19 &  5.24 &  2.86 & 91.90 & \textbf{93.33} \\
      GPT o4-mini   & 10 s  & T/F   & S-\rev{hop} &  3.33 &  5.00 & 91.67 & \textbf{94.17} &  0.56 &  0.56 & 98.89 & \textbf{99.17} \\
      GPT o4-mini   & 10 s  & T/F   & M-\rev{hop} & 14.29 & 12.14 & 73.57 & 79.64 &  1.78 &  3.56 & 94.67 & \textbf{96.44} \\
      \cmidrule(lr){2-12}
      GPT o4-mini   & 100 s & Expl. & S-\rev{hop} & 12.50 & 12.50 & 75.00 & \textbf{81.25} &  6.15 &  6.15 & 87.69 & \textbf{90.76} \\
      GPT o4-mini   & 100 s & Expl. & M-\rev{hop} & 25.64 & 17.95 & 56.41 & 65.38 &  7.69 & 12.82 & 79.49 & \textbf{85.90} \\
      GPT o4-mini   & 100 s & MC    & S-\rev{hop} & 17.39 & 11.59 & 71.01 & 76.81 &  4.96 &  2.48 & 92.56 & \textbf{94.05} \\
      GPT o4-mini   & 100 s & MC    & M-\rev{hop} & 22.22 & 11.11 & 66.67 & 72.22 &  5.88 &  2.94 & 91.18 & \textbf{92.65} \\
      GPT o4-mini   & 100 s & T/F   & S-\rev{hop} &  6.58 &  7.89 & 85.53 & \textbf{89.47} &  1.54 &  1.92 & 96.54 & \textbf{97.50} \\
      GPT o4-mini   & 100 s & T/F   & M-\rev{hop} & 18.25 & 13.14 & 68.61 & 75.18 &  2.55 &  3.63 & 93.83 & \textbf{95.64} \\
      \bottomrule
    \end{tabular}%
  }
\end{table}

To better understand how user's query style and graph query structure interact, we grouped all model responses by question category and visualized the average accuracy in each bucket. Specifically, we aggregate scores across all three LLM models, then compute mean accuracy for every combination of \emph{user question type} and \emph{graph question type}. Figure~\ref{fig:rq4_heatmap} summarizes the result of this experiment. Adding KG in TAAF improves accuracy in every cell. The largest gain appears in True/False multi-\rev{hop} queries, which rise by +30.12 percentage to reach 79.63\%. Explanatory questions remain the most challenging across the board, but TAAF still boosts their accuracy by 23.89\% in baseline setting. Looking at the absolute values, the highest observed accuracy is 90.99\% for True/False single-\rev{hop} queries under the TAAF setting.
On the other hand, explanatory multi-\rev{hop} queries are the lowest at 37.22\% under the Baseline setting, reflecting the compound reasoning required. A full breakdown by model and configuration is provided in Table~\ref{tab:combined_results}.

\medskip
\begin{flushleft}
\begin{tikzpicture}
\node[
  draw=blue!70!black,
  fill=blue!8,
  thick,
  rounded corners=6pt,
  inner sep=7pt,
  anchor=west,
  text width=\dimexpr\linewidth-15pt\relax
] {
  \textbf{RQ1 Key observations.}
  \begin{itemize}[leftmargin=*]
    \item True/False single-\rev{hop} questions achieve the highest accuracy overall (90.99\%).
    \item Multi-\rev{hop} queries are harder in general but benefit more from graph grounding.
    \item Explanatory prompts remain the most difficult, especially on multi-\rev{hop} questions, despite strong gains.
  \end{itemize}
};
\end{tikzpicture}
\end{flushleft}

\FloatBarrier
\textbf{RQ2: What is the effect of time-interval length on the performance of TAAF's answers?}

\begin{figure}[H]
  \centering
  \includegraphics[width=.9\linewidth]{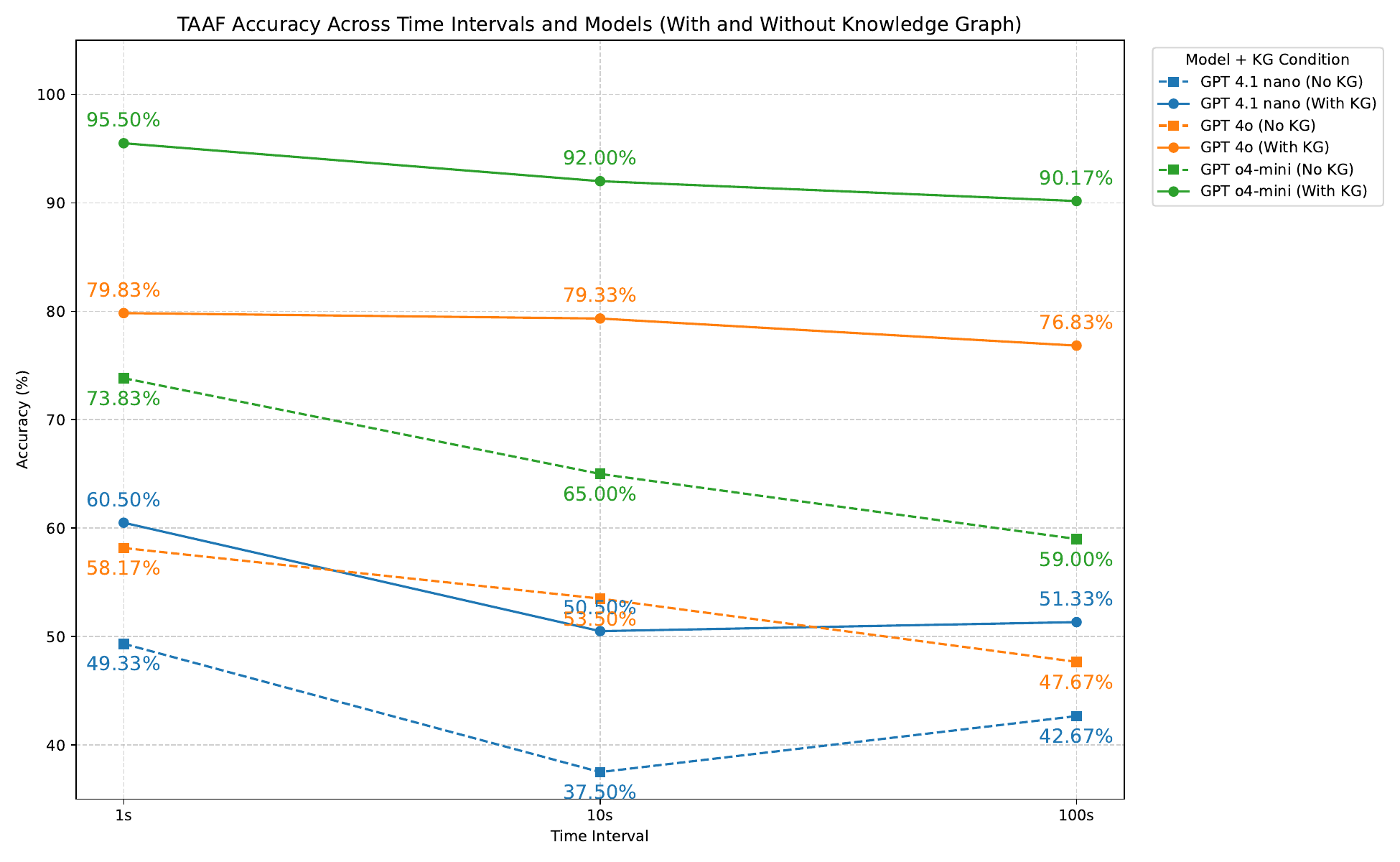}
  \caption{ \revm{Accuracy at 1s/10s/100s for each model, with (solid) vs. without (dashed) KG}{Accuracy at for each model and interval}}
  \label{fig:rq2_models}
\end{figure}

Next, we examine how the time interval length affects performance.
To isolate the effect of time-interval length, we contrasted the three time windows 1s, 10s, and 100s across all three LLM settings with and without the knowledge graph (TAAF vs. Baseline).
Figure~\ref{fig:rq2_models} plots the resulting accuracy curves.  Solid markers denote the TAAF variant and dashed markers trace the baseline. The results (TAAF) reveal a clear, model-agnostic trend: accuracy falls as the window widens, but the rate of decline depends on model strength.
\medskip
\noindent\textbf{Numerical drop‐off (TAAF Setting).}
\begin{itemize}[leftmargin=*]
  \item \textbf{o4-mini} 95.5\,\% → 92.0\,\% → 90.17\,\%  
       (–5.33 pp overall)
  \item \textbf{GPT-4o} 79.83\,\% → 79.33\,\% → 76.83\,\%  
        (–3.0 pp)
  \item \textbf{GPT-4.1 nano} 60.5\,\% → 50.5\,\% → 51.33\,\%  
        (–9.17 pp)
\end{itemize}
\noindent\textbf{Numerical drop-off (baseline setting).}
\begin{itemize}[leftmargin=*]
\item \textbf{o4-mini} 73.83 \% → 65.0 \% → 59.0 \% (–14.83 pp overall)
\item \textbf{GPT-4o} 58.17 \% → 53.5 \% → 47.67 \% (–10.5 pp)
\item \textbf{GPT-4.1 nano} 49.33 \% → 37.5 \% → 42.67 \% (–6.66 pp)
\end{itemize}

Two forces compete here. Longer windows supply richer relational evidence, yet they also enlarge the State System output or the graph. Stronger models like o4-mini and GPT-4o exploit the extra context with only modest degradation, whereas the lightweight GPT-4.1 nano loses grounding signal more rapidly. The baseline decline is steeper because it must process ever-larger volumes of raw State System data, while TAAF mostly needs to update node and edge features as the graph grows. We also observe two small upticks in the blue GPT-4.1 nano curves; one with KG on (50.5 → 51.33\%) and one in the baseline (37.5 → 42.67\%) when moving from the 10s to the 100s window. While the exact cause remains speculative and needs further investigation, one plausible explanation is that the wider temporal window enables the model to access more cumulative evidence, which benefits some of queries that depend on aggregated behavior. Additionally, GPT-4.1 nano may struggle with the intermediate 10s input which is too complex to reason over fully but not large enough to highlight dominant patterns. 

\medskip
\begin{flushleft}
\begin{tikzpicture}
\node[
  draw=blue!70!black,
  fill=blue!8,
  thick,
  rounded corners=6pt,
  inner sep=7pt,
  anchor=west,
  text width=\dimexpr\linewidth-15pt\relax
] {
  \textbf{RQ2 Key observations.}
  \begin{itemize}[leftmargin=*]
    \item \emph{Shorter windows are the safest.}  
          All models peak on the 1-second slice; noise accumulation
          outweighs extra context beyond ~10s.
    \item \emph{o4-mini remains robust.}  
          Even at 100s it retains 90\%+ accuracy, suggesting that careful reasoning objectives can handle graph expansion.
    \item \emph{Baseline deteriorates faster.}  
        In baseline settings, accuracy drops roughly two-to-three times as quickly as in TAAF, underscoring how raw State-System output becomes unwieldy as the interval grows.
  \end{itemize}
};
\end{tikzpicture}
\end{flushleft}

\FloatBarrier

\textbf{RQ3: To what extent does the incorporation of a Knowledge Graph improve the accuracy of the TAAF when answering trace-related queries?}
\label{sec:rq1_accuracy_gain}

\begin{figure}[H]
  \centering
  \includegraphics[width=.9\linewidth]{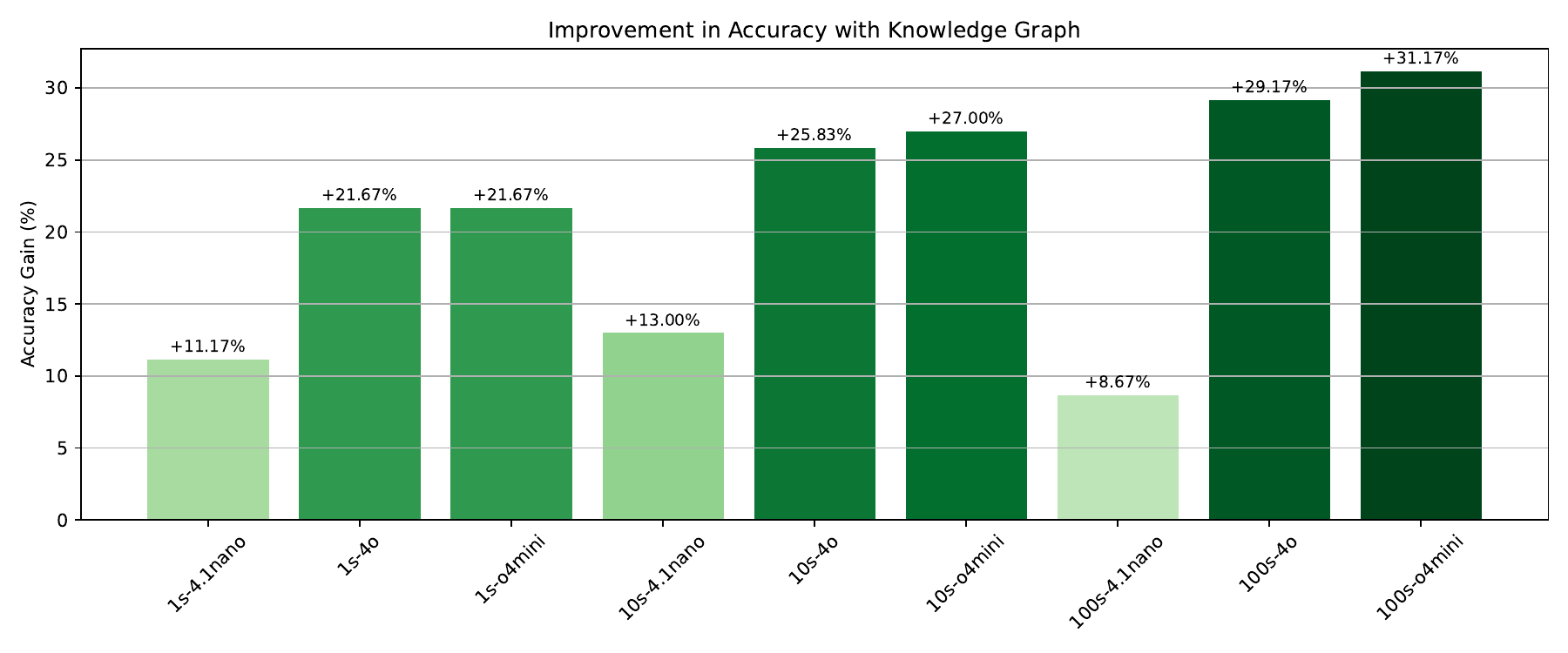}
  \caption{Accuracy gain from KG in TAAF \revm{(\% over baseline) by model and interval}{}.}
  \label{fig:rq1_gain}
\end{figure}

Third, we quantify how much explicit structure within KG helps over purely numeric State System data. Figure~\ref{fig:rq1_gain} quantifies the absolute accuracy lift obtained by adding
the knowledge-graph context in TAAF compared with the baseline. Across all nine model–interval combinations the KG consistently helps, ranging from a modest \(+\,8.67\%\) for \texttt{GPT-4.1 nano} on the 100-second slice up to a sizeable \(+\,31.17\%\) for \texttt{o4-mini} on the 100-second slice. The mean gain over the entire grid is \(21.5\%\), confirming that graph grounding is the most influential factor in our pipeline.

\begin{figure}[H]
  \centering
  \includegraphics[width=.9\linewidth]{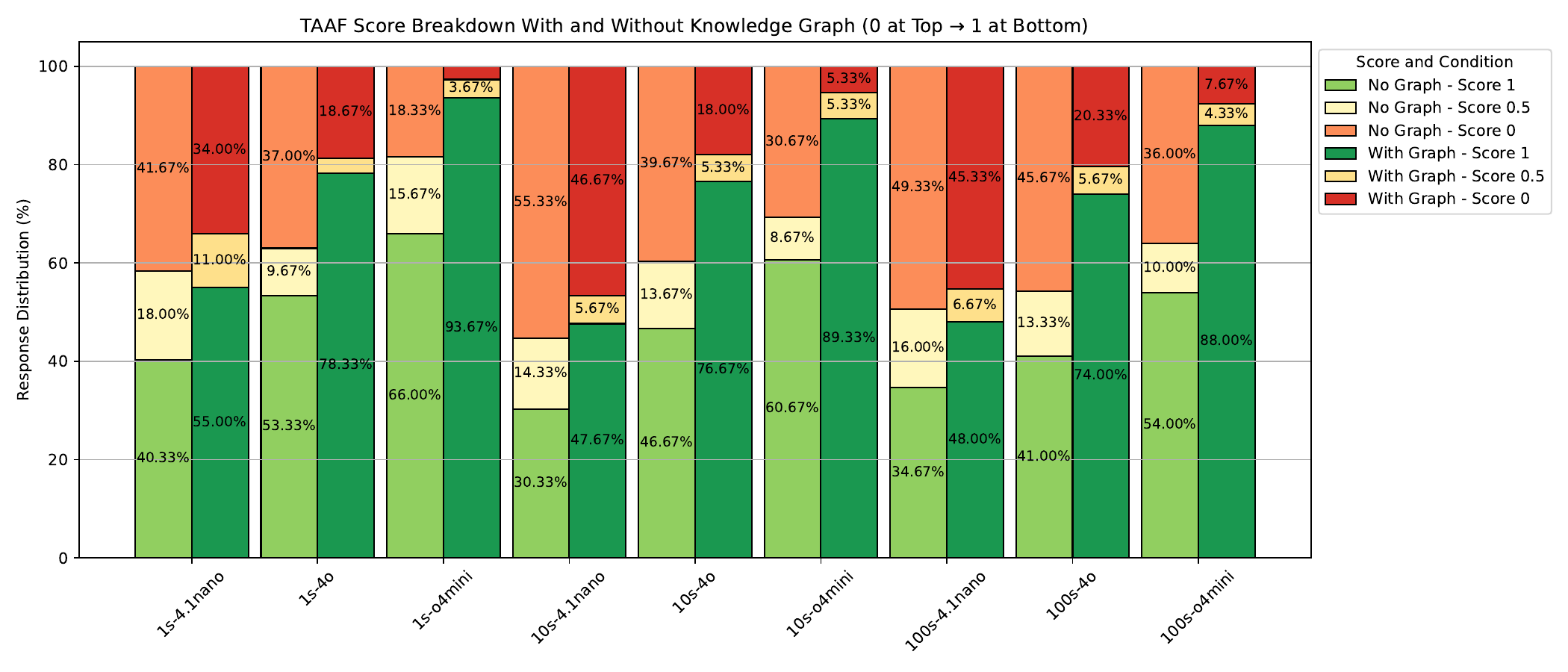}
  \caption{Score breakdown (0/0.5/1) with vs. without KG. \revm{taller green = more correct}{} }
  \label{fig:rq1_breakdown}
\end{figure}

Figure~\ref{fig:rq1_breakdown} explains how the accuracy rises in more details. For every model–interval pair the left stacked bar is the baseline and the right stacked bar is the TAAF variant. The KG chiefly converts partial or wrong answers into fully correct ones. The fraction of 0.5-scores also falls, indicating the model is not merely guessing and it reaches a decisive, correct conclusion more often.

\medskip
\noindent
\begin{tikzpicture}
\node[
  draw=blue!70!black,
  fill=blue!8,
  thick,
  rounded corners=6pt,
  inner sep=7pt,
  anchor=west,
  text width=\dimexpr\linewidth-15pt\relax 
] (box) at (0,0) {
  \textbf{RQ3 Key observations.} 
  \begin{itemize}[leftmargin=*]
    \item \emph{Longer intervals benefit more.}
          The lift grows steadily from 1s to 100s because multi-hop relations become denser and harder to reconstruct from raw numerics alone.
    \item \emph{Stronger models still improve.}
          Even the larger models (\texttt{GPT-4o}) gain roughly
          \(+\,22\%\) at 1s and \(+\,29\%\) at 100s, showing that
          parametric knowledge does not replace explicit structure.
    \item \emph{Reasoning-tuned models exploit the KG best.}
          \texttt{o4-mini} shows the highest incremental benefit,
          suggesting that lightweight reasoning objectives align
          well with graph-grounded setting.
  \end{itemize}
};
\end{tikzpicture}

\textbf{RQ4: How does the accuracy and quality of TAAF responses vary across different LLM backends \rev{(including a focused cross-family probe)}?}

Finally within Phase 1, we ask whether the observed gains depend on the underlying LLM. Because the same experiment grid spans all three LLMs, we can read the impact of interval length directly from Figure~\ref{fig:rq2_models}.

Overall, o4-mini with TAAF settings tops the chart at every interval, peaking at 95.5\% on the 1-second slice and still clearing 90.17\% on the 100-second slice.  GPT-4o trails by roughly 15\%, while the compact GPT-4.1 nano stays in the 50–60\% band even with graph support.  With baseline settings the spread narrows, confirming that stronger reasoning models capitalize on explicit structure more fully than weaker ones.

\begin{figure}[H]
  \centering
  \includegraphics[width=\linewidth,height=.32\textheight,keepaspectratio]{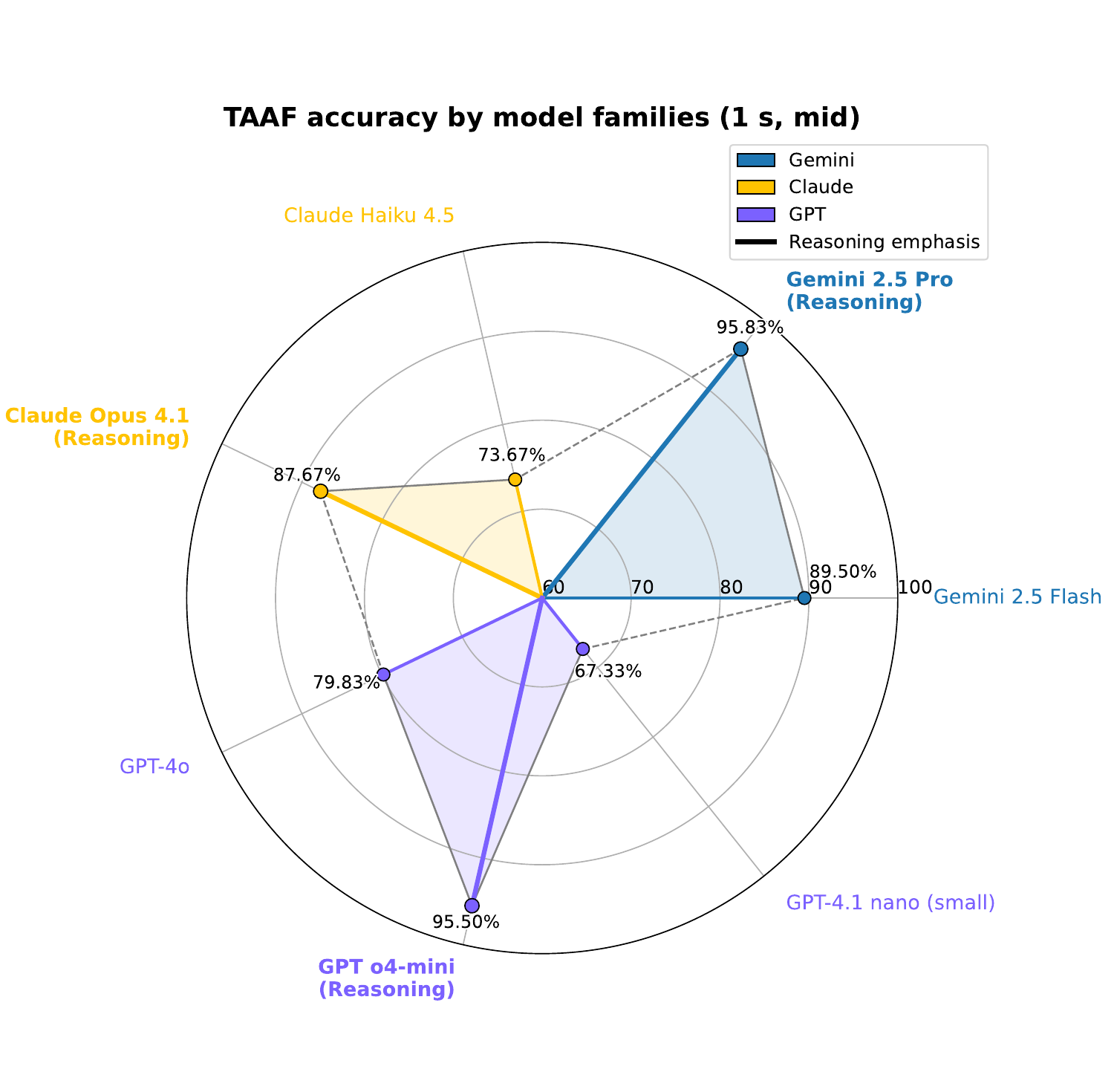}
  \caption{\rev{TAAF accuracy across model families.}}
  \label{fig:rq4_cross_family}
\end{figure}

\rev{\paragraph{Cross-family probe.}
To check that TAAF's accuracy and findings do not depend on a single vendor family, we also ran a focused probe on two additional model families under the same prompt and scoring protocol. We selected one flagship and one reasoning-tuned model per family to mirror the GPT tiers. The probe uses the \textbf{TAAF} setting on a \textbf{1\,s mid-trace} slice to bound runtime and isolate the model family factor. Accuracies are Gemini 2.5 Pro (Reasoning) \textbf{95.83\%}, Gemini 2.5 Flash \textbf{89.50\%}, Claude Opus 4.1 (Reasoning) \textbf{87.67\%}, Claude Haiku 4.5 \textbf{73.67\%}. The pattern mirrors the GPT tiering. Reasoning-oriented flagships top their family, while compact fast models trade a modest drop for speed. Fig.~\ref{fig:rq4_cross_family} better presents the numbers and indicates that the accuracy comes from the TAAF graph grounding rather than vendor specifics. This provides evidence that TAAF carries across model families and supports general use.}

\medskip
\begin{flushleft}
\begin{tikzpicture}
\node[
  draw=blue!70!black,
  fill=blue!8,
  thick,
  rounded corners=6pt,
  inner sep=7pt,
  anchor=west,
  text width=\dimexpr\linewidth-15pt\relax
] (box) at (0,0) {
  \textbf{RQ4 Key observations.}
  \begin{itemize}[leftmargin=*]
    \item \emph{o4-mini with TAAF settings leads.}  
          Its accuracy exceeds 90\% at every interval. This
          shows the value of explicit reasoning objectives when graph context is available.
    \item \emph{Gap widens with structure.}  
          The performance delta between GPT-4o and GPT-4.1 nano grows
          once the KG is supplied, suggesting larger models extract
          richer relational cues.
    \item \rev{\emph{Vendor-agnostic pattern.}
          The cross-family probe mirrors the GPT tiering.
          Gaps stay within a single-digit band when tiers are matched, which points to TAAF’s graph grounding as the main source of the lift.}
  \end{itemize}
};
\end{tikzpicture}
\end{flushleft}

\subsection{Phase 2: Focused Factors}
\textbf{RQ5: Does providing the LLM with the graph schema (node types and features) enhance the accuracy of its responses?}

Having established the KG’s value, we test whether giving the model a lightweight JSON schema improves grounding further. Before running the full suite of experiments, we first asked whether the LLM actually \emph{uses} the structural hints contained in the schema (node types and edge features). We therefore repeated the GPT-4o mid-trace, 1-second experiment with the schema strings removed from the prompt but kept the graph triples intact.  Table~\ref{tab:rq5_schema} shows the outcome.

\begin{table}[H]
  \centering
  \caption{Impact of passing the graph schema.}
  \label{tab:rq5_schema}
  \small
  \begin{tabular}{lcccc}
    \toprule
    \textbf{Prompt Variant} & \textbf{0} & \textbf{0.5} & \textbf{1} & \textbf{TAAF Accuracy} \\
    \midrule
    No schema & 20.0 \% & 16.7 \% & 63.3 \% & 71.7 \% \\
    With schema & 18.7 \% & 3.0 \% & 78.3 \% & \textbf{79.8 \%}\,\textcolor{green!60!black}{\scriptsize$\uparrow$} \\
    \bottomrule
  \end{tabular}
\end{table}

\medskip
\begin{flushleft}
\begin{tikzpicture}
\node[
  draw=blue!70!black,
  fill=blue!8,
  thick,
  rounded corners=6pt,
  inner sep=7pt,
  anchor=west,
  text width=\dimexpr\linewidth-15pt\relax
] (box) at (0,0) {
  \textbf{RQ5 Key observation.}
  Passing the lightweight schema yields an extra
  \(+\,8.1\%\) absolute accuracy; therefore all remaining
  experiments keep the schema enabled.
};
\end{tikzpicture}
\end{flushleft}

\textbf{RQ6: How does the temporal location, that is the position of the time window within the trace, affect TAAF accuracy?}

\begin{figure}[H]
  \centering
  \includegraphics[width=.9\linewidth]{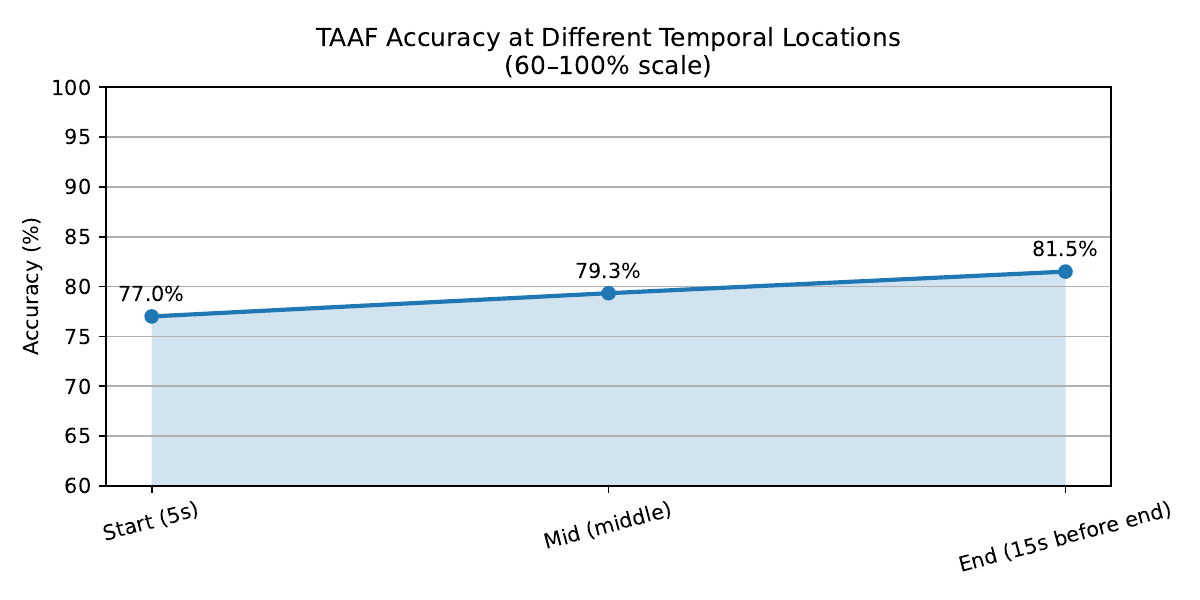}
  \caption{TAAF accuracy \revm{when the analysis window is placed at the start, middle, or end of the trace.}at different temporal locations}
  \label{fig:rq6_temporal}
\end{figure}

Timing matters: we now check if accuracy shifts when the time window is taken early, mid-way, or late in the trace. For RQ6 we fixed GPT-4o in the TAAF setting and applied a 10-second analysis window at three temporal positions.
We tested three placements of that window in each trace.  
The \textbf{start} location covers from 5 seconds into the trace until 15 seconds in. The \textbf{mid} location spans from the midpoint of each trace to 10 seconds beyond that midpoint.  
The \textbf{end} location covers from 15 seconds before the trace ends until 5 seconds before the end.  
Figure~\ref{fig:rq6_temporal} shows the details of this experiment.

 As shown in Figure \ref{fig:rq6_temporal} accuracy climbs from 77.0\% at the start, to 79.3\% in the middle, and reaches 81.5\% near the end.  
Later windows often contain longer, steadier bursts of activity, making scheduling patterns easier to spot.  
Early windows cover the warm-up phase of the workload and include more transient behaviour.  
The overall spread is small, about 4\%, so users can place the window almost anywhere in the trace and still receive reliable answers.

\medskip
\begin{flushleft}
\begin{tikzpicture}
\node[
  draw=blue!70!black,
  fill=blue!8,
  thick,
  rounded corners=6pt,
  inner sep=7pt,
  anchor=west,
  text width=\dimexpr\linewidth-15pt\relax
] {
  \textbf{RQ6 Key observations.}
  \begin{itemize}[leftmargin=*]
    \item Accuracy slightly improves as the window moves from the start toward the end, plus 4.5\% overall.
    \item Late segments feature steadier execution phases, giving clearer patterns for the model to follow.
    \item The small gap shows that TAAF remains robust to window placement, so users need not fine-tune the timestamp.
  \end{itemize}
};
\end{tikzpicture}
\end{flushleft}

\textbf{RQ7: How does the sampling temperature influence both accuracy and consistency?}
Finally, we explore the sensitivity of both accuracy and stability to the LLM’s sampling temperature.

\begin{figure}[H]
  \centering
  \includegraphics[width=.9\linewidth]{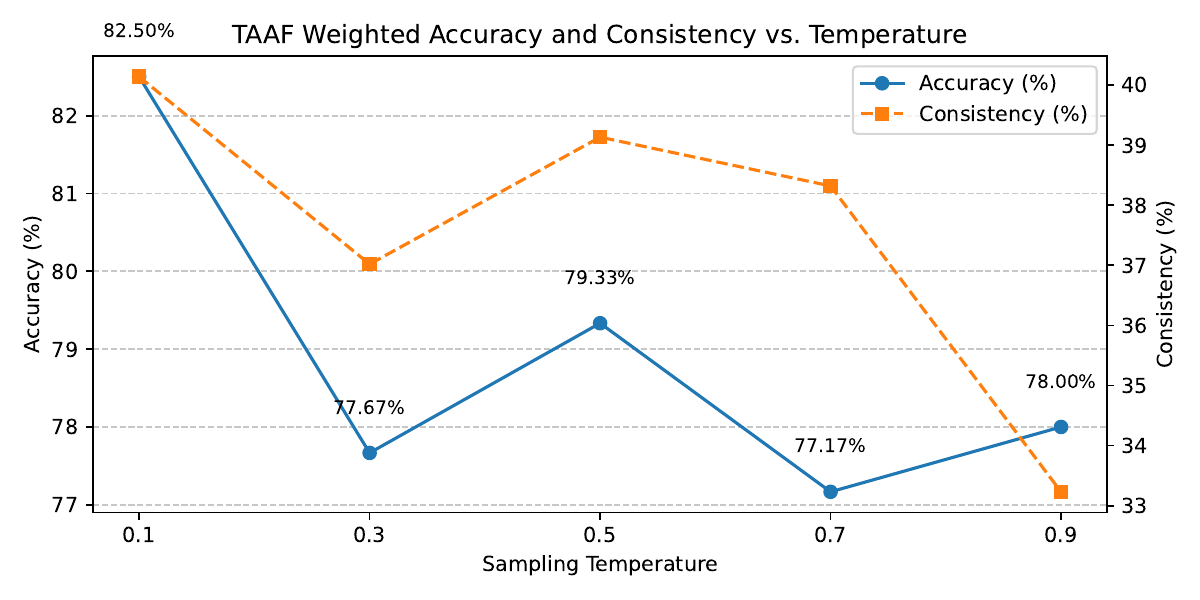}
  \caption{Accuracy and consistency as a function of sampling temperature.}
  \label{fig:rq7_temp}
\end{figure}

Like in RQ6, we kept GPT-4o in TAAF setting and the 10 second mid-trace window, then varied the sampling temperature that controls randomness in the LLM output.  
Five values were tested: 0.1, 0.3, 0.5, 0.7, and 0.9.  
For each value we computed accuracy and the consistency metric defined in Section \ref{sec:evaluation-setup}.  
Figure \ref{fig:rq7_temp} plots the results.

Based on the results, accuracy is highest at the lowest temperature, 82.5\% at 0.1, then drops to 77.67\% at 0.3, recovers slightly at 0.5, and declines again at 0.7. Consistency follows a similar but smoother curve: best at 0.1, stable near 0.5, and lowest at 0.9. Low temperatures reduce hallucination but risk deterministic errors, while very high temperatures introduce too much noise.

\medskip
\begin{flushleft}
\begin{tikzpicture}
\node[
  draw=blue!70!black,
  fill=blue!8,
  thick,
  rounded corners=6pt,
  inner sep=7pt,
  anchor=west,
  text width=\dimexpr\linewidth-15pt\relax
] {
  \textbf{RQ7 Key observations.}
  \begin{itemize}[leftmargin=*]
    \item Best trade-off appears at 0.1 to 0.3, high accuracy with the top consistency scores.
    \item Temperatures above 0.7 lower accuracy and consistency, showing the model drifts when sampling is too random.
    \item A moderate setting near 0.5 balances diversity and reliability, which is why earlier experiments used this as the default.
  \end{itemize}
};
\end{tikzpicture}
\end{flushleft}

\subsection{Discussion}
\label{sec:Discussion}

The two-phase evaluation shows that TAAF raises accuracy in every
setting while remaining stable across models, window sizes, and query
styles.  Graph grounding supplies explicit structure, short windows
limit noise, and a lightweight schema boosts grounding at little extra
cost.  These results justify the design choices made in
Section~\ref{sec:methodlogy}. The State System gives fast,
time-indexed access, the query-specific knowledge graph keeps the
context small and relevant, and the LLM converts that context into
actionable insight.

We have also highlighted where performance still falls, such as
explanatory multi-\rev{hop} questions and very long windows.  These gaps
suggest clear avenues for future optimization and help frame the limitations that follow.

With the empirical evidence in place, we now examine possible threats to the validity of our study.

\section{Threats to Validity}
\label{sec:validity}

We outline potential limitations that may affect the interpretation of our results and suggest directions for future improvement.

\subsection{Construct Validity}

\textbf{Question diversity.}  
Our benchmark includes 100 hand-crafted questions from SciMark 2.0 traces. While diverse in format and scope, it may miss edge cases like rapid context switching or deeply nested interrupt chains. Future work can extend coverage via automated generation.

\textbf{Scoring resolution.}  
Our 3-level rubric \{0, 0.5, 1\} captures broad correctness but cannot distinguish minor errors (e.g., off-by-one times). More fine-grained metrics (e.g., BLEU, numeric tolerance) could improve sensitivity.

\subsection{Internal Validity}

\textbf{Annotation bias.}  
Author-labeling may bias scores, especially on borderline outputs. Using multiple independent raters and reporting inter-annotator agreement would increase objectivity.

\textbf{Prompt length limits.}  
Some large traces or graphs exceed LLM context limits. We truncate edge attributes as needed, which may reduce answer quality. Future solutions include hierarchical prompting or sliding windows.

\subsection{External Validity}

\textbf{Trace generalizability.}  
All questions are based on SciMark 2.0 under Linux. Other kernels or workloads could alter behavior. Broader evaluation using diverse traces (e.g., eBPF, HPC, mobile) is needed.

\textbf{Model stability.}  
LLMs accessed via APIs may evolve silently, affecting reproducibility. Caching model versions and outputs can help mitigate this threat.

\rev{\textbf{Scope limits.} TAAF is most effective when queries are scoped to a window. Global, unbounded queries can produce KGs too large for contemporary contexts or collapse temporal order if summarized aggressively. For such cases, global pre-aggregation or approximate sampling is more appropriate.}

\rev{\textbf{LLM-as-reasoner limits.} We observed even with a correct KG context, models sometimes fail on multi-step aggregation. For example, the query “which CPU has the lowest difference between its highest and lowest accumulated times among threads” requires multiple aggregation steps. We observed that models may skip steps or make small arithmetic slips.}

\subsection{Conclusion Validity}

\textbf{Sampling variability.}  
Each query was sampled three times. While standard practice, higher replication or paired tests would improve statistical confidence.

\textbf{Baseline scope.}  
We compare TAAF only to the raw State System. Intermediate designs, such as summaries or static graphs, could further contextualize TAAF's gains.



\section{Conclusion and Future Work}
\label{sec:conclusion}

This paper introduced \textbf{TAAF}, a framework that transforms low-level kernel traces into query-relevant answers by integrating temporal indexing, knowledge graphs, and large language models. TAAF addresses key challenges in trace analysis, including volume, temporal complexity, semantic ambiguity, and the need for domain expertise, by combining \revm{symbolic}{} abstractions with structured context for reasoning.

Across 7,800 responses spanning three \revm{LLMs}{GPT models}, three time windows, and six query types, graph-based grounding improved weighted accuracy by an average of 21.5\%, reaching up to 95.5\% on short intervals. These gains confirm that injecting explicit structure enhances both factuality and stability, even for high-performing models.

Evaluation also revealed that longer windows and multi-hop queries remain challenging, though reasoning-oriented models like o4-mini remain robust. TAAF's performance is stable across early, mid, and late trace slices, and sampling temperatures between 0.1 and 0.3 offer a strong trade-off between accuracy and diversity.

\revm{Looking forward, promising directions include integrating temporal knowledge graphs for continuous reasoning, embedding TAAF into autonomous agent workflows, applying the framework to production-scale systems to surface real bugs, and scaling beyond current context limits through retrieval-augmented or hierarchical prompting.}{Looking forward, promising directions include integrating temporal knowledge graphs for continuous reasoning and embedding TAAF into autonomous agent workflows. Additional work will apply the framework to production-scale systems to surface real bugs and scale beyond current context limits through retrieval-augmented or hierarchical prompting. A complementary direction is a systematic latency review that profiles each pipeline stage from end to end while exploring optimizations such as graph partitioning and sampling, streaming or incremental KG construction, batching, and early-exit heuristics.}



\bibliographystyle{ACM-Reference-Format}
\balance
\bibliography{references}

\end{document}